\documentclass[]{aa}
\usepackage{graphicx}
\usepackage{txfonts}
\usepackage{xcolor}
\usepackage{graphicx} \graphicspath{ {./Images/GJ357b/} } \usepackage{amsmath}
\usepackage{amssymb}
\usepackage{multicol}
\usepackage[export]{adjustbox}
\usepackage[flushleft]{threeparttable}
\begin{document}

   \title{GJ~357~b}

   \subtitle{A Super-Earth Orbiting an Extremely Inactive Host Star}

   \author{D. Modirrousta-Galian\inst{1,2} \and B. Stelzer\inst{3,2} \and E. Magaudda\inst{3} \and J. Maldonado \inst{2} \and M. G\"udel \inst{4} \and J. Sanz-Forcada\inst{5} \and B. Edwards\inst{6} \and G. Micela\inst{2} }

   \institute{Universit\`a di Palermo, Department of Physics and Chemistry, Via Archirafi 36, 90126 Palermo, Italy
   	\and
   	INAF – Osservatorio Astronomico di Palermo, Piazza del Parlamento 1, I-90134 Palermo, Italy
         \and
         Institut f\"ur Astronomie \& Astrophysik, Eberhard-Karls Universit\"at T\"ubingen, Sand 1, D-72076 T\"ubingen, Germany
         \and
         Institut f\"ur Astrophysik, Universit\"at Wien, T\"urkenschanzstrasse 17, A-1180 Wien, Austria
         \and 
         Departamento de Astrofísica, Centro de Astrobiología (CSIC-INTA), ESAC Campus, Camino bajo del Castillo s/n, E-28692 Villanueva de la Cañada, Madrid, Spain
         \and
         Department of Physics and Astronomy, University College London, London, United Kingdom
         }
         
   \date{}

  \abstract
   {}
   {In this paper we present a deep X-ray observation of the nearby M dwarf GJ\,357 and use it to put constraints on the atmospheric evolution of its planet, GJ\,357\,b. We also analyse the systematic errors in the stellar parameters of GJ~357 in order to see how they affect the perceived planetary properties.}
   {By comparing the observed X-ray luminosity of its host star, we estimate the age of GJ\,357\,b as derived from a recent {\em XMM-Newton} observation {($\log{L_{\rm x}}\,{\rm [erg/s]} = 25.73$), with $L_{\rm x} -$ age relations for M dwarfs. We find that GJ\,357 presents one of the lowest X-ray activity levels ever measured for an M dwarf, and we put a lower limit on its age of $5$\,Gyr.} Using this age limit, we performed a backwards reconstruction of the original primordial atmospheric reservoir. Furthermore, by considering the systematic errors in the stellar parameters, we find a range of possible planetary masses, radii, and densities.}
   {From the backwards reconstruction of the irradiation history of GJ\,357\,b's we find that the upper limit of its initial primordial atmospheric mass is $\sim \rm 38M_{\oplus}$. An initial atmospheric reservoir significantly larger than this may have survived through the X-ray and ultraviolet irradiation history, which would not be consistent with current observations that suggest a telluric composition. However, given the relatively small mass of GJ~357~b, even accreting a primordial envelope $\gtrsim 10~\rm M_{\oplus}$ would have been improbable as an unusually low protoplanetary disc opacity, large-scale migration, and a weak interior luminosity would have been required. For this reason, we discard the possibility that GJ~357~b was born as a Neptunian- or Jovian-sized body. In spite of the unlikelihood of a currently existing primordial envelope, volcanism and outgassing may have contributed to a secondary atmosphere. Under this assumption, we present three different synthetic IR spectra for GJ\,357\,b that one might expect, consisting of $100\%~\rm CO_{2}$, $100\%~\rm SO_{2}$, and  $75\%~ \rm N_{2}$, $24\%~\rm CO_{2}$ and $1\%~\rm H_{2}O$, respectively. Future observations with space-based IR spectroscopy missions will be able to test these models. Finally, we show that the uncertainties in the stellar and planetary quantities do not have a significant effect on the estimated mass or radius of GJ~357~b.}
   {}


   \maketitle
%
%
\section{Introduction}

	Nearby transiting planets, especially those also detected with the
	radial-velocity method, are very important targets for complete
	characterisation. A subset of these interesting bodies are terrestrial planets
	with masses comparable to that of the Earth. Even though theoretical models predict that
	they are very abundant \citep[e.g.][]{Schlichting2013,Simon2016}, they are rare
	within our astronomical catalogues due to instrumental limitations. Therefore,
	detailed analyses of these planets are valuable in order for us to understand
	their physical properties. From this perspective, GJ\,357\,b is an intriguing
	exoplanet. It orbits a nearby star ($\pi = 105.88 \, mas \pm 0.06$ from Gaia
	DR2) that displays very low magnetic activity. The optical emission line activity of GJ\,357 is 
	among the lowest for its spectral type \citep{Schoefer2019}, 
	the Ca\,{\sc ii} $\log{R_{\rm HK}^\prime}$ value is $-5.37$ \citep{BoroSaikia2018},
	and a photometric rotation period of $\sim 78$\,d was inferred from combining
	data from different ground-based surveys \citep{Luque2019}. The planet, GJ\,357\,b,
	was detected by TESS (TOI 562.01) in Sector 8, with a transit depth of
	$1164 \pm 66$~ppm and a periodicity of $3.93$~days, corresponding to an orbital distance of $0.035 \pm 0.0002$ AU. The system also includes two non-transiting super-Earths in wider orbits which appear to have larger masses \citep[$M \sin{i} = 3.40 \pm 0.46\,{\rm M_\oplus}$ and $M \sin{i} = 6.1 \pm
	1\,{\rm M_\oplus}$ for GJ\,357\,c and GJ\,357\,d respectively;][]{Luque2019}.

	Regarding GJ~357~b, its mass and radius measurements are $1.84 \pm 0.31\,M_\oplus$ and $1.22 \pm 0.08\,R_\oplus$ (i.e. a density of $\rm 5.6 \pm 1.2~\rm g~cm^{-3}$ ), respectively
	\citep{Luque2019}, which is consistent with a telluric planet that most probably
	does not host a primordial atmosphere. Given that it may host a secondary atmosphere, and considering that secondary atmospheres can provide important clues as to the internal structure and evolution of planets, we believe
	GJ\,357\,b is one of the best targets for spectroscopic observations with future
	instrumentation such as the James Webb Space Telescope (JWST, \cite{greene}), the Extremely Large Telescope (ELT, \cite{brandl_ELT}), Twinkle \citep{twinkle}, and Ariel \citep{tinetti_ariel}. It is therefore important to predict what kind of atmosphere could exist on this planet and the type of evolution that it has undergone throughout its lifetime.
	
	In this paper we evaluate the evolution and composition of the atmosphere of GJ\,357\,b based on the age estimate derived from our measurement of the X-ray luminosity of the host star, GJ\,357. In Sect.~\ref{sect:xrays} we present the new {\em XMM-Newton} data, their analysis, and interpretation including the age estimate we derive. In Sect.~\ref{sect:backwards_reconstruction}
	we use that result to show that GJ~357~b may have formed with a primordial atmosphere as large as $\sim 38~\rm M_{\oplus}$. However, we then argue that an atmosphere of this size would have been unlikely due to requiring an atypical formation history. After this, we produce synthetic spectra in Sect.~\ref{sec:synthetic_atmosphere} to show what one might expect from observational data of its atmosphere. In Sect.~\ref{sect:stellarparams} we revisit and compare different ways to estimate the stellar parameters in order to assess the influence of these values on the planet radius and mass. We discuss and summarise our results in Sect.~\ref{sect:discussion}.
	
\section{Constraints from X-ray data}\label{sect:xrays}

    GJ\,357 was observed with {\em XMM-Newton} on 19 May 2019 (Obs-ID\,0840841501) in the course of a systematic survey of X-ray activity in nearby M dwarfs (PI Stelzer). These observations were designed to reach the deep sensitivity required to constrain the very faint X-ray luminosities of the least active M dwarf stars. Specifically, the X-ray properties derived from such data, in combination with the rotation period and in comparison to other samples of M dwarfs, enable us to estimate the age of the star, which is relevant for an assessment of the evolution of its planets.  

\subsection{X-ray data analysis}\label{subsect:xrays_analysis}

    We analysed the X-ray observation with the {\em XMM-Newton} 
	Science Analysis System (SAS)\footnote{SAS Data Analysis Threads: 
	\url{https://www.cosmos.esa.int/web/xmm-newton/sas-threads}} 18.0 pipeline. 
	Our analysis is focused on  the data from the EPIC/pn instrument which provides the highest
	sensitivity. 
	
	The first analysis step was the extraction 
	of the light curve for the whole detector which we inspected for time-intervals of
	high background. We identified such intervals using as a cutoff threshold a count rate 
	value of $0.35$\,cts/s. Those parts of the observations with a count rate above this
	value were excluded from the subsequent analysis, while the remaining ones define the 
	good time intervals (GTIs). We found this method to give a relatively clean observation and a nominal exposure time of 29.76\,ks which is reduced only by a few percent to 29.0\,ks of GTIs. In addition, we filtered the data for pixel pattern (PATTERN$\leq$12) for an optimum trade-off between detection efficiency, spectral resolution, and quality flag (FLAG=0); and we only retained events with energy greater than $0.2$\,keV where the spectral response function is well calibrated.  
	
	We performed the source detection using the SAS pipeline {\sc edetect\_chain} 
		in the full EPIC/pn energy range ($0.15-12.0$\,keV) and in five narrower bands: 
		(1) $0.15-0.3$\,keV, (2) $0.3-1.0$\,keV, (3) $1.0-2.4$\,keV, (4) $2.4-5.0$\,keV, 
		(5) $5.0-12.0$\,keV. This analysis showed that above $5.0$\,keV the count rate of GJ\,357 
		is approximately zero. The number of net source counts in the $0.15-5.0$\,keV band 
		(i.e. the sum of bands 1 ...4) is $118$\,cts.   
	
	The extraction of spectrum and light curve was performed considering a
	 source region of $15''$ centred on the position obtained for GJ\,357 from the
	 source-detection process, with an annulus for the 
	 background region centred at the same position and with an inner radius three 
	 times larger than the source radius. We created the response matrix and ancillary 
	 response for the spectral analysis with the SAS tools, 
	 \textit{\sc rmfgen} and \textit{\sc arfgen}, and we grouped the
	 spectrum to a minimum of five counts for each 
	 background-subtracted spectral channel. 
	 We initially extracted the spectrum of GJ\,357 in the $0.15-5.0$\,keV band, as suggested by the source detection. We then noticed that below $0.3$\,keV the source shows count fluctuations that do not follow the mean spectral trend. For this reason, the spectral fitting within the XSPEC environment\footnote{XSPEC NASA's HEASARC Software:\url{https://heasarc.gsfc.nasa.gov/xanadu/xspec/}} was performed for the $0.3-5.0$\,keV band. First, we used a one-temperature APEC model with frozen abundances ($\rm Z = 0.3\,Z_{\odot}$). We then included a new model component (cflux) to calculate the flux in the chosen energy band, keeping the APEC model components fixed to the values obtained in the first step. The best-fitting model yields the coronal temperature ($kT$), the emission measure ($EM$), and the flux ($f_{\rm x}$);  all listed in Table~\ref{APEC_results}. Due to the low number of counts in the spectrum this simple spectral shape provides an appropriate description of the data. We also tried to vary the abundances, and to fix them on the solar value, but this led to significantly poorer fits ($\chi_{\rm red}^2 \sim 2$).
	 
	 We extracted the light curve in the same energy band as the spectrum ($0.3-5.0$\,keV) using the SAS pipeline {\sc epiclccorr} which applies the background subtraction, and the corrections for dead time, exposure, and GTI. The light curve does not present obvious evidence for variability such as flares. 
   	
	\begin{table}[t]
	\caption{X-ray spectral parameters of GJ\,357 and flux in the $0.3-5.0$\,keV band.}            
	\begin{center}
	    \begin{threeparttable}
		\begin{tabular}{ccccc}
			\hline
			$kT$&$\log{EM}$ & $\chi_{\rm red}^2 $ & d.o.f.$^{\ast}$ &  $f_{\rm x}\,(0.3-5.0\,\rm keV)$ \\
			${\rm [keV]}$ & [${\rm cm^{-3}}$] & & & [${\rm 10^{-15} ergs/cm^2/s}$] \\
			\hline
			&&&&\\
			0.20$\pm$0.10 & 47.69$\pm$0.16 & 0.9 & 13 & $(2.90 \pm 0.88)$\\
			\hline
		\end{tabular}
		\begin{tablenotes}
        \small
        \item The 2~$\sigma$ uncertainties were computed with the error pipeline provided in the XSPEC software package. $^{\ast}$short for degrees of freedom.
        \end{tablenotes}
        \end{threeparttable}
		\label{APEC_results}
	\end{center}
\end{table}

	The majority of historical X-ray measurements for M dwarfs refer to the $0.1-2.4$\,keV band. Therefore, in order to be consistent with data from the literature to which we make reference in Sect.~\ref{subsect:xrays_discussion} we estimated the flux in the $0.1-2.4$\,keV energy band with XSPEC for our best-fitting model from Table~\ref{APEC_results}, obtaining a scaling factor between the two energy bands (from $0.3-5.0$\,keV to $0.1-2.4$\,keV) amounting to $1.73\pm0.73$. We then calculated the X-ray luminosity in the $0.1-2.4$\,keV band from the scaled flux and the {\em Gaia} distance (9.44$\pm$0.01\,pc) that we calculated from the parallax given in the {\em Gaia-DR2} archive. Because it is known that {\em Gaia-DR2} contains spurious astrometric solutions \citep{Arenou2018}, we verified that the parallax of GJ\,357 is reliable by evaluating the quality flags provided by \citet{Lindegren2018}. The resulting value for the $0.1-2.4$\,keV X-ray luminosity is $\log{L_{\rm x}}\,{\rm [erg/s]} = 25.73\pm0.23$. The relative error includes the uncertainty of the flux in the $0.3-5.0$\,keV energy band weighted with that of the scaling factor. We also used the {\em Gaia} distance together with the 2MASS apparent $K_{\rm s}-$magnitude ($K_{\rm s} = 6.47 \pm 0.02$\,mag) to compute the absolute magnitude in the $K_{\rm s}-$band ($M_{\rm K_s}$). We combined $M_{\rm K_s}$ with the solar bolometric magnitude ($M_{\rm bol,\odot}  = 4.75$) to calculate the bolometric 
	luminosity of GJ\,357 ($L_{\rm bol} = 0.015 \pm 0.001\,{\rm L_\odot}$). Using this value we found a fractional X-ray luminosity of $\log{(L_{\rm x}/L_{\rm bol})} = -6.05\pm0.09$.

\subsection{The X-ray activity of GJ\,357 in context and an age estimate}\label{subsect:xrays_discussion}

It is instructive to compare the X-ray properties of GJ\,357 to those of other M dwarfs. Particularly relevant in the context of our article are the clues that X-ray activity can give on stellar age. We therefore compare GJ\,357 to the recent results of \cite{Magaudda2020}. The sample of this latter study includes both new data and a comprehensive compilation of the previous literature on the X-ray activity$-$rotation$-$age relation of M dwarfs. \cite{Magaudda2020} analysed all data in a homogeneous way, which required some updates to the literature results
including the use of Gaia parallaxes.

\begin{figure*}
\begin{multicols}{2}
\includegraphics[width=0.5\textwidth]{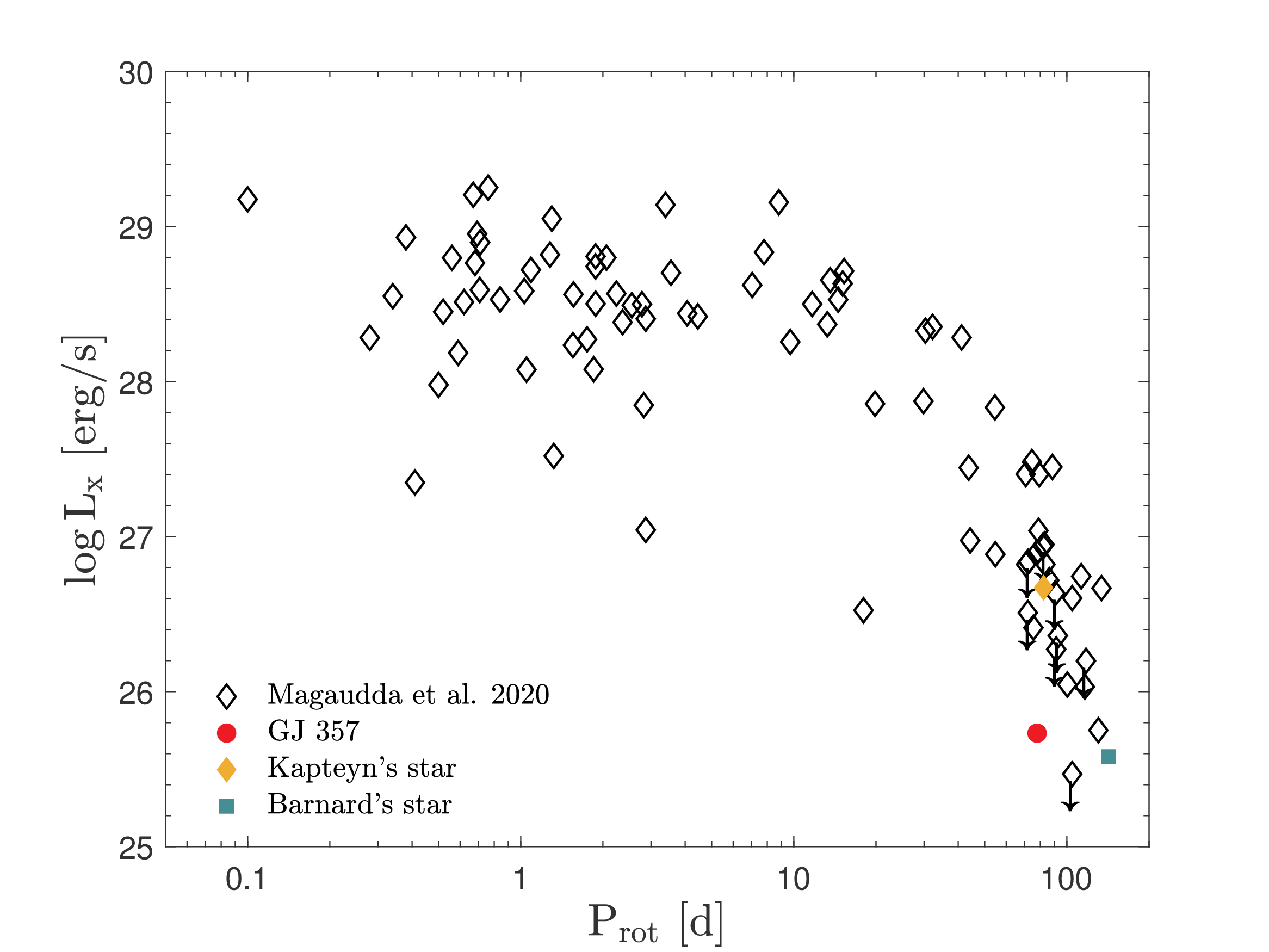}\par 
\includegraphics[width=0.5\textwidth]{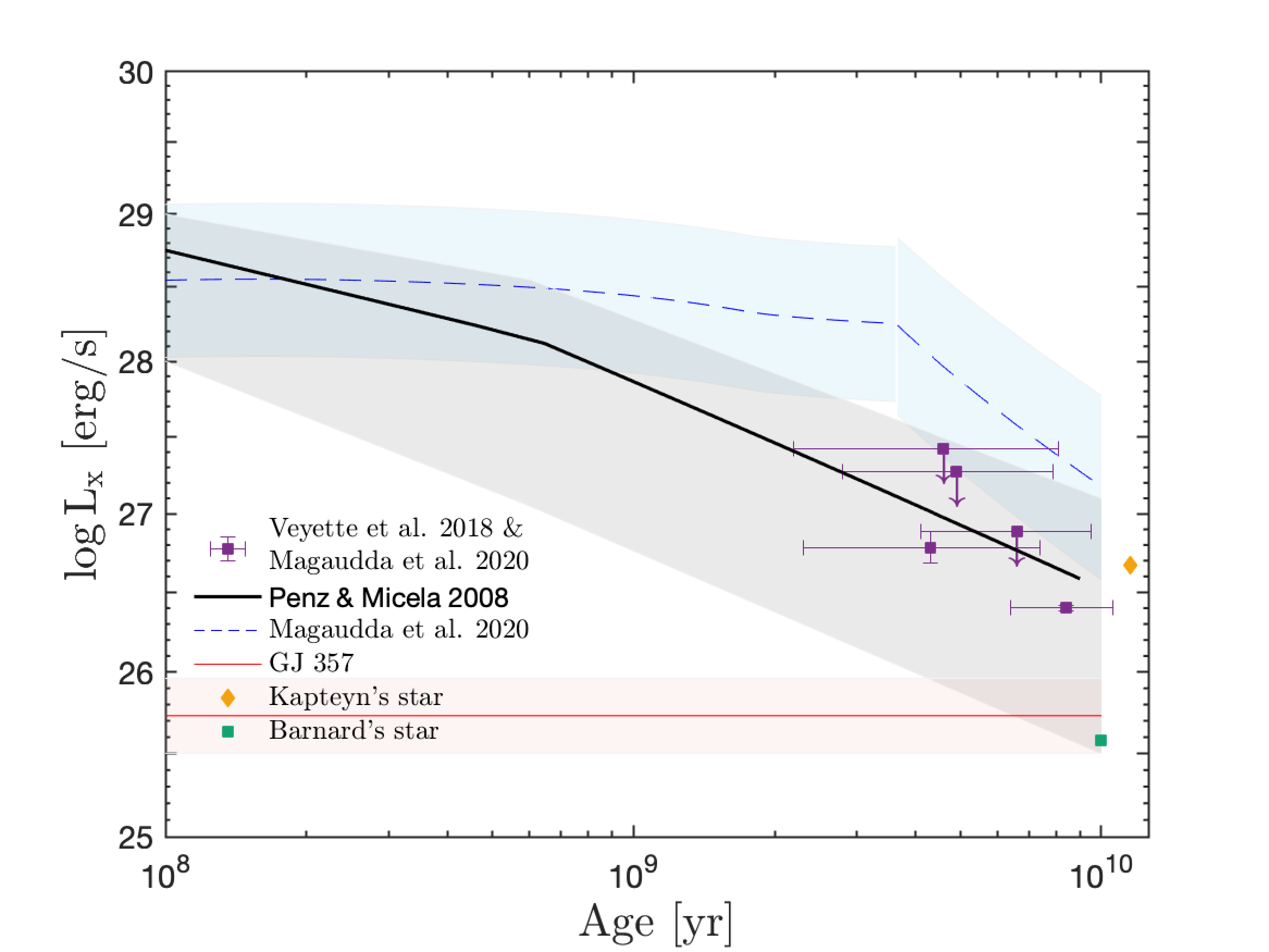}\par 
\end{multicols}
\caption{GJ\,357 compared to the activity$-$rotation$-$age relation of M dwarfs from 
\protect\cite{Magaudda2020}. {\em Left}: X-ray luminosity vs. rotation period; for GJ\,357 the $L_{\rm x}$ measurement is derived from recent {\em XMM-Newton} data analysed in this paper and the $P_{\rm rot}$ value is from \protect\cite{Luque2019}. {\em Right}: X-ray luminosity vs. age reconstructed based on the observed $L_{\rm x} - P_{\rm rot}$ relation (left panel) and the angular momentum evolution models of \cite{Matt2015}. In addition, the empirical scaling law from \cite{Penz2008(1)} is shown. Two M stars with known ages, Barnard's star (GJ~699) and Kapteyn's star (GJ~191),  are added to the plots for reference.}
\label{fig:xrays_in_context} 
\end{figure*}

In Fig.~\ref{fig:xrays_in_context} we present the relation between X-ray luminosity and rotation period ($L_{\rm x}$ vs $P_{\rm rot}$) 
and the $L_{\rm x}-$age relation from \cite{Magaudda2020} together with the respective parameter values of GJ\,357. The sample shown here is restricted to stars in the mass range $0.14 - 0.4\,{\rm M_\odot}$, to which GJ\,357 belongs according to the stellar mass
presented by \cite{Luque2019} ($0.34\,{\rm M_\odot}$) and our own analysis of the stellar parameters presented in Sect.~\ref{sect:stellarparams} of this paper. Remarkably, the X-ray luminosity of GJ\,357 is at the same level as that of the least active and slowest rotating M dwarfs in the same mass range (Fig.~\ref{fig:xrays_in_context} - left panel). Only the exceptional sensitivity of {\em XMM-Newton} allows us to probe such low activity levels of even the most nearby M dwarfs. 

The right hand side of Fig.~\ref{fig:xrays_in_context} shows a derived $L_{\rm x}-$age relation based on the observed $L_{\rm x} - P_{\rm rot}$ relation (left panel) and the angular momentum evolution models ($P_{\rm rot} - \rm age$) from \cite{Matt2015}, where the light blue shade represents the uncertainty of the X-ray luminosity at a given age, mainly originating from the spread of $L_{\rm x}$ at a given rotation period. Ages for field M dwarfs are
extremely hard to determine and are available only for a handful of stars. In Fig.~\ref{fig:xrays_in_context} we overplot stars in the examined mass range $0.14-0.4\,{\rm M_\odot}$ 
that have known individual ages. This sample includes some field M dwarfs, with ages derived by \cite{Veyette2018} from kinematics and chemical evolution and X-ray luminosities presented by \cite{Magaudda2020}. Moreover, we have included the notorious benchmark M dwarfs Kapteyn's star and Barnard's star with X-ray luminosities from \cite{Schmitt1995}, rotation periods from \cite{Guinan2016} and \cite{ToledoPadron2019}, and ages from \cite{Guinan2016} and \cite{Riedel2005}.

As can be seen from Fig.~\ref{fig:xrays_in_context} (right panel) the field stars with known age have systematically fainter X-ray activity than predicted by the $L_{\rm x} -$age relation from \cite{Magaudda2020}. This disagreement between observed and predicted $L_{\rm x} -$age relations is not seen in the higher mass M and K dwarfs ($0.4-0.6\,{\rm M_\odot}$ and $0.6-0.8\,{\rm M_\odot}$ bins) that were studied in an analogous way by \cite{Magaudda2020}. The likely cause of this discrepancy in the lowest-mass M dwarfs are shortcomings in the angular evolution models, which fail to predict the very long rotation periods of these objects, resulting in an over-prediction of their X-ray luminosities \citep[see][for further details]{Magaudda2020}. We have overplotted in Fig.~\ref{fig:xrays_in_context} (right) the empirical scaling relation $L_{\rm x}(t)$ from \cite{Penz2008(1)} which is based on observed X-ray luminosity functions (XLFs) of the Pleiades, the Hyades, and of nearby field stars. The X-ray luminosity function of the field M stars was derived from \citet{Schmitt2004}  where stars were selected mainly from a modified version of the CNS\,3 \citep[short for the \textit{Catalog of Nearby Stars};][]{Gliese1991}, with the addition of nearby, very late-type stars discovered by near-infrared surveys. In order to estimate the decay of $L_X$ with age, and because the determination of individual age for dM stars is very difficult, \cite{Penz2008(1)} assigned an average age of $6$\,Gyr to sample. Furthermore, these latter authors assumed that the luminosity of Prox Cen is a good estimate of the mean value of the X-ray luminosity function at $6$\,Gyr and that the observed spread is representative of the spread at each age (shown as the grey area around the relation in the figure). The final scaling law was obtained  by interpolating between the data of the Pleiades, the Hyades, and the field stars.

While Fig.~\ref{fig:xrays_in_context} shows that the shape of the $L_{\rm x}-$age relation of M dwarfs remains uncertain, it nevertheless allow us to put a constraint on the age of GJ\,357 directly from its observed X-ray luminosity (shown by the horizontal line in the right panel of Fig.~\ref{fig:xrays_in_context} with its uncertainty in light-red shading). Similar to the other field M dwarfs, GJ\,357 lies well below the X-ray luminosity for stars of its mass predicted by \cite{Magaudda2020}. Moreover, the $L_{\rm x}$ of GJ\,357 is lower than that of all M dwarfs with known age except for Barnard's star and is marginally consistent with the faint end of the dispersion around the  distribution of \cite{Penz2008(1)}.

We conclude that the age of GJ\,357 is most likely
higher than $\sim 5$\,Gyr.

\section{Backwards reconstruction}
\label{sect:backwards_reconstruction}

    It is common within the literature to perform backwards reconstructions of exoplanet atmospheres in order to determine their histories \citep[e.g.][]{Locci2019,Modirrousta2020}. This is important as it can provide us with useful information concerning the formation and evolution of exoplanets. For instance, the bimodal distribution of exoplanet radii is believed by many to be caused by photoevaporation, which has been scrutinised by performing backwards reconstructions of exoplanet populations \citep[e.g.][]{Owen2013,Owen2017,Modirrousta2020b}. Furthermore, the remnant cores of planets, which once hosted large primordial atmospheres, are of great interest to the exoplanetary community as they provide us with an inference for the interior structure of bodies like Jupiter and Saturn \citep[e.g.][]{Mocquet2014,toi849}. Accordingly, it is useful to know the size of the primordial atmosphere of GJ~357~b in order to better understand its formation, evolution, and current properties, and to decipher whether or not a star with an unusually low XUV flux can still greatly influence its host planets.
    
    Before one can progress with the backwards reconstruction, there are a few aspects that need to be considered. The first is that given the mass and radius of this body, a hydrogen-rich atmosphere is unlikely. However, the observed density could be the result of a very iron-rich embryo that is engulfed within a thin primordial envelope. There are three main formation scenarios that could lead to a super-ferruginous composition. The first concerns surface vaporisation due to high temperatures. Whilst this may have taken place on Mercury \citep{Cameron1985}, for more massive planets the efficiency of this mechanism drops precipitously due to the increasing gravitational force (Ito et al. in prep). In addition, for the extreme thermodynamic conditions required for this process to take place, the retention of a hydrogen atmosphere is incompatible. Another possibility is that GJ~357~b formed exotically due to chemical/thermal gradients within the protoplanetary disc \citep{Lewis1972,Lewis1974}. The third potential argument is that collisional stripping resulted in the silicate mantle layers becoming ejected which results in a relatively high iron abundance. This mechanism has also been proposed for Mercury \citep{Benz1988} and has been adopted to explain the density trends of some exoplanets \citep{Marcus2010,Swain2019}. 
	
\begin{figure}[ht] 
	\centering
	\includegraphics[scale=0.55]{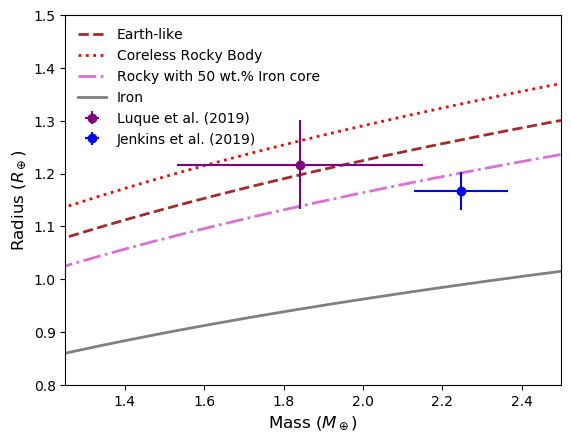} 
	\caption{Mass vs. radius for GJ 357 b using the coreless rocky (red dotted line), Earth-like (dark red dashed line), rocky with a $\rm 50~wt.\%$ iron core (purple dash-dot line), and pure iron (grey line) models from \citet{Zeng2013} and \citet{Zeng2016}. The values for the mass and radius and their uncertainties are those from \citet[][purple]{Luque2019} and \citet[][blue]{Jenkins2019}.}
	\centering 
	\label{fig:mass_radius} 
	\end{figure}

	However, these ‘exotic’ scenarios are unlikely as they require atypical formational paths and a relatively tenuous hydrogen atmosphere which is sensitive to atmospheric escape and would therefore have a short lifespan at the current orbital distance of GJ\,357\,b \citep[e.g.][]{Sanz-Forcada2011,Ehrenreich2011, Lammer2013, Owen2013, Jin2014, Owen2017, Jin2018, Kubyshkina2018(2),Locci2019,Modirrousta2020b}. The atmospheric lifespan would be short because an M star is particularly active in its first year of life and so heavy mass losses would be expected. Therefore, while this peculiar composition is possible, GJ~357~b is most probably of Earth-like composition (see Fig.~\ref{fig:mass_radius}). 
	
    Given that the age of GJ\,357\,b (assumed to be coeval with its host star) is $\rm > 5$\,Gyr, and that it most probably does not presently host a hydrogen envelope (see Sect.~\ref{sect:backwards_reconstruction}), one cannot accurately determine when its primordial atmosphere was lost, if it had one at all. This is problematic, because if one assumes that the atmosphere was lost earlier than it actually was, the primordial atmospheric mass would be underestimated. On the other hand, if the assumption for the time of the complete atmospheric loss is later than the true value then the reservoir would be overestimated. Furthermore, having a mass of $\rm 1.84 \pm 0.31M_{\oplus}$ \citep[and a radius of $\rm 1.217 \pm 0.084R_{\oplus}$;][]{Luque2019} the planets' gravity may have been too weak to accrete a large atmosphere. However, one could argue that if the surrounding nebular gas has a sufficiently small grain opacity and the planet has a low internal luminosity, enough nebular gas could be accreted to form a Neptunian- or even Jovian-sized body \citep{Ikoma2000} even for a relatively small body like GJ~357~b.
    
    To test this assumption, we performed a backwards reconstruction in which we determined the upper bound of GJ\,357\,b's initial atmospheric mass. This is the maximum primordial atmospheric mass that would have been fully eroded by XUV evaporation. This implies that an initial atmospheric reservoir smaller than or equal to this critical mass would be consistent with the current telluric composition. To do this we adopted the hydro-based approximation for XUV-induced evaporation of \citet{Kubyshkina2018(1)}. We started from the derived age of the planet system ($\simeq\rm 5~Gyr$) and we linearly extrapolated the XUV luminosity of the host star backwards in time. This was done by using the lower bound of the X-ray luminosity predicted by the scaling law from \citet{Penz2008(1)}, which overlaps with the uncertainty range of the observed X-ray luminosity of GJ~357. Considering the lower bound rather than the average X-ray$-$age relation, we account for the unusually low X-ray luminosity of GJ~357  (which can be seen in Fig.~\ref{fig:xrays_in_context}). Our simulation ended when the star's age was $10^{7}$ years and its X-ray luminosity was $\rm \approx 10^{29}~erg~s^{-1}$. The starting value of $L_{\rm x}$ obtained this way is in sufficient agreement with the observed range of X-ray luminosities for low-mass stars in the $10$\,Myr-old TW\,Hya association ($10^{29}$ ... $10^{30}\,{\rm erg~s^{-1}}$) derived by \citet[][see our comment at the end of the following paragraph]{Kastner2016}.

    From the backwards-calculated X-ray luminosity history of GJ\,357 we estimated its EUV luminosities across time ($L_{\rm EUV}$) by adopting the scaling law between $L_{\rm EUV}$ and $L_{\rm x}$ given by \citet{Sanz-Forcada2011}. We input these values into the analytical `hydro-based' evaporation model by \citet{Kubyshkina2018(1)}. In addition, for the radius evolution of the planet we adopted the mass$-$radius relation of a silicate embryo engulfed by a hydrogen envelope from \citet{Lopez2014}. In a similar manner to these latter authors, we used the photosphere as the cut-off point for the planetary radius. Furthermore, their model only works for planets older than $100~\rm Myr$ whereas we begin photoevaporation at $10~\rm Myr$. In order to account for this, we assumed a constant interior luminosity from $10$ to $100~\rm Myr$. However, we did not incorporate planetary migration into our mass-loss model \citep[see][for an example of evaporation-induced planetary migration]{Modirrousta2020}. Therefore, by estimating how much hydrogen was lost from the moment the planet was born, we constrained the maximum allowed initial atmospheric reservoir. We note that the unknown details of the shape of the age decay of the stellar X-ray emission (see Sect.~\ref{subsect:xrays_discussion}) do not affect the result considerably because most of the atmospheric erosion occurs in a very short time compared with the stellar age of this system. Furthermore, our result is robust against different detailed backwards paths in the $L_X - \rm age$ plane, since dM stars of the mass of GJ 357 evolve close to the saturation regime at least for 1 Gyr. Therefore, even if we were incorrect by an order of magnitude in estimating the precise value of the X-ray luminosity at a given time, the planet would still have lost its primordial atmosphere.
   
    Our approach provides a theoretical upper bound on the initial atmospheric mass of $\rm \sim 38M_{\oplus}$ with a radius of $\rm \sim 14R_{\oplus}$. This is an upper bound because an atmospheric mass smaller than $\rm \sim 38M_{\oplus}$ would have also resulted in the currently observed density, which is consistent with a telluric composition. The uncertainties in these predicted values are quite large, especially as the mass$-$radius relations for silicate bodies with hydrogen-rich envelopes are not yet fully understood. For example, if one assumes a high abundance of radioactive species then the internal luminosity would have resulted in a significantly larger radius than if one adopts a less luminous embryo \citep{Lopez2014}. In addition, the available literature shows a large variation in the radius of hydrogen-rich planets for low masses \citep[e.g.][]{Lissauer2011,Rogers2011,Lopez2012,Lopez2014}, which will strongly influence the mass-loss history.
    
    In any case, the large initial atmospheric mass predicted by our backwards reconstruction shows that even for stars that currently have remarkably low XUV luminosities, the mass-loss effects on their host planets can be substantial. Nevertheless, considering the small mass of GJ~357~b, accreting a hydrogen envelope of $\rm \sim 38M_{\oplus}$ may have been difficult as an unusually low protoplanetary disc grain opacity, a very low embryo luminosity, and large-scale migration would have been required. This is improbable as firstly the metallicity of the star is only slightly below that of the Sun (see sect.~\ref{sect:stellarparams}), from which we can infer that the protoplanetary disc had a relatively normal grain opacity. Secondly, in order to achieve low internal luminosities, GJ~357~b must have either lost the majority of its formational energy prior to accreting its hydrogen envelope \citep[e.g.][]{Ikoma2000}, or stored it efficiently \citep[e.g.][]{Jespersen2020} and released it slowly over time. Whilst both of these modes are possible, generally it is believed that the embryo's luminosity plays a strong role in inhibiting gas accretion and even triggering its evaporation \citep{Ginzburg2016,Ginzburg2018}. Thirdly, had GJ~357~b formed far out and then migrated inwards, one would expect a lower bulk density consistent with icy species, which is not compatible with the observed mass and radius measurements \citep{Zeng2013,Zeng2016,Zeng2018}. If one instead adopts a more typical accretionary model (i.e. with standard grain opacities, core luminosities, and a smaller migration) this would lead to a primordial atmosphere $\rm \lesssim 0.02M_{\oplus}$ \citep[calculated using the models from][]{Ikoma2012,Chachan2018} which would have been fully eroded away by XUV-irradiation in less than $\rm 10~Myr$ \citep{Kubyshkina2018(1)}. Considering all of the above, our calculation shows that GJ~357~b was probably not born as a Neptunian body or a gas giant which contrasts strongly with planets such as 55~Cancri~e or CoRoT-7b that had the potential to have been significantly more massive in the past due to their large masses and tight orbits \citep[e.g.][]{Ehrenreich2011,Kubyshkina2018(2)}.
   
    Conversely, \citet{Jenkins2019} find that GJ~357~b has a substantially higher density, consistent with a metal-rich composition (i.e. a core that is $\simeq 60~\rm wt.\%$ of the total mass). However, an alternative explanation for this high density is that GJ~357~b is a compressed, icy remnant core of a planet that originally had a large hydrogen envelope \citep{Mocquet2014, Modirrousta2020b}. New observational data further support this possibility as shown by the `T2' trend in \citet{Swain2019}. The label `T2' is given to a population of planets which increase in density for progressively smaller radii. \citet{Swain2019} propose that this trend could be caused by highly compressed remnant cores. The mass and radius values of GJ~357~b from \citet{Jenkins2019} lie within the density trend found by \citet{Swain2019}, which could be suggestive of a highly compressed planet that once hosted a large hydrogen envelope \footnote{\citet{Swain2019} used the energy-limited mass-loss equation \citep[e.g.][]{Erkaev2007} which usually predicts mass losses $100-1000$ (and sometimes as much as $\sim 10^{9}$) times smaller than the hydro-based model, which includes thermal effects \citep{Kubyshkina2018(1)}. Therefore, even though the authors focus on highly irradiated planets, the greater influence that stars have on super-Earths and sub-Neptunes predicted by the hydro-based approximation shows that the `T2' trend is also relevant to our situation.}. Comparing the results from \citet{Luque2019} and \citet{Jenkins2019} shows that there is a clear ambiguity in the composition and history of GJ~357~b. However, it must be noted that there is a small overlap in the uncertainties of \citet{Luque2019} and \citet{Jenkins2019} that allows for both mass and radius values to be equal and compatible with a more typical rocky composition with a $\simeq 50~\rm wt.\%$ iron core. Nevertheless, a spectroscopic analysis of the atmospheric composition could be performed to reduce this ambiguity, and we discuss this possibility further in Sect.~\ref{sec:synthetic_atmosphere}.

\section{Synthetic atmospheric spectra}
\label{sec:synthetic_atmosphere}
    
    Whilst GJ~357~b probably lost its primordial atmosphere, geological processes may have subsequently formed a secondary, volcanic one. Therefore, performing a spectroscopic analysis may reveal insights into its internal structure, the redox state of the mantle, and its history.
    
    Several studies investigate the outgassing of super-Earths \citep[e.g.][]{Kite2009,Noack2017,Dorn2018}. The interior modelling of super-Earths and sub-Neptunes is a deeply complex issue, and a thorough review cannot be made within this manuscript. Nevertheless, within the literature it is generally believed that when silicate bodies become more massive than $\sim 5~\rm M_{\oplus}$, convection becomes inhibited \citep[e.g.][]{Tackley2013,Miyagoshi2015,Miyagoshi2017,Miyagoshi2018, Dorn2018}. However, GJ~357~b is small enough that most models are consistent with convection and hence volcanism being plentiful. For instance, if one assumes a stagnant lid model then \citet{Dorn2018} predict an outgassed mass of $\sim 10^{20}~\rm kg$ ($\rm \sim 5 \times 10^{-5}~M_{\oplus}$) for $\rm CO_{2}$. Notwithstanding, other typical volcanic gases such as $\rm SO_{2}$ and $\rm H_{2}O$ would also be expected, although the redox state of the mantle will ultimately dictate the chemical composition of the atmosphere. For example, if GJ~357~b formed under oxygen-poor conditions, one might expect more reduced atmospheric species such as $\rm H_{2}S$. These distinct models could be verified with a spectroscopic analysis of the atmosphere which will be possible with future space missions (explained later in this section).
    
    In this section we use the TauREx code \citep{Alrefaie2019} to generate synthetic spectral models for GJ\,357\,b based on our assumption that a volcanic atmosphere is present and that GJ\,357\,b did not form under reduced conditions. TauREx is a Bayesian program that is optimised to process the molecular line lists from the ExoMol project \citep[see][for details]{Tennyson2016} to either generate forward spectral models of exoplanetary atmospheres or, in its retrieval mode (inverse model), interpret exoplanet atmospheric data by fitting them with a transmission or emission model. To create a forward spectrum, the following parameters are required: 
	\begin{itemize} 
	\item Temperature and spectral type of the host star. 
	\item Mass ($M_{\rm p}$) and radius ($R_{\rm p}$) of the exoplanet. 
	\item Atmospheric thermodynamic properties. 
	\item Gas mixing ratios. 
	\item Wavelength range modelled. 
	\item Presence of clouds or hazes and their properties
	(i.e. distribution, location, particle size and shape). 
	\end{itemize} 
	For the stellar temperature, spectral type, planetary radius, and planetary mass we used the values given in \citet{Luque2019} which are $\approx 3505~\rm K$, M2.5~V, {\rm $1.84\,{\rm R_\oplus}$}, and {\rm $1.217\,{\rm M_\oplus}$} respectively\footnote{In sect.~\ref{sect:stellarparams} we present alternative methods to obtain the stellar mass and radius and the values derived with them.}. 
	For the stellar spectra we use the data files from the PHOENIX library \citep[for more information see][]{Hauschildt1999,Hauschildt2010}.
	
	To maintain consistency with the simplicity of our models, we adopt an isothermal atmospheric profile within the photospheric region. This is a reasonable assumption as the effective temperature of GJ~357~b is large ($> 500~\rm K$). We simulate the atmosphere using the plane-parallel approximation, with pressures ranging from $10^{-9}$ to $10^2$ bar, uniformly sampled in log-space with 100 atmospheric layers. We note that the actual surface pressure is unknown as it is very sensitive to the specific atmospheric properties such as the opacities, composition, and temperature profile. However, at pressures above 10 bar, the atmosphere is likely opaque and therefore does not contribute to any of the observed features. Here, we present three potential compositions: $100\%~\rm CO_{2}$, $100\%~\rm SO_{2}$, and $\rm 75\%~ N_{2}$, $\rm 24\%~ CO_{2}$ and $\rm 1\%~ H_{2}O$. 
	
\begin{figure}[ht]
	\centering
	\includegraphics[width=\columnwidth]{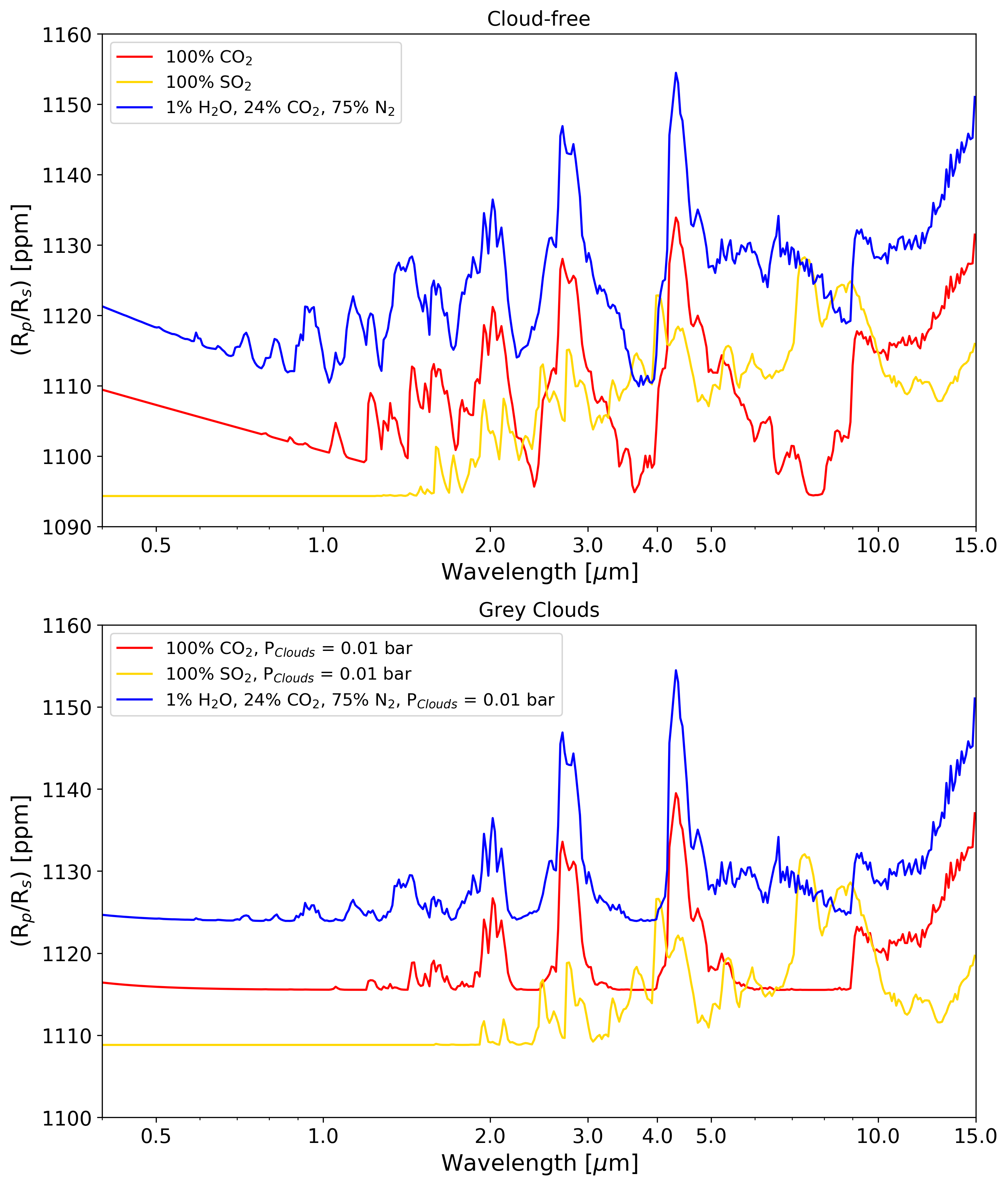}
	\caption{Synthetic spectra of GJ~357~b generated using the TauREx code. Our diagram includes the synthetic spectra of $100\%~\rm CO_{2}$ (red), $100\%~\rm SO_{2}$ (yellow), and a mixed atmosphere with $75\%~\rm N_{2}$ along with $1\%~\rm H_{2}O$ and $24\%~\rm CO_{2}$ (blue). We show these forward models for the cloud-free case (top) and in the presence of opaque grey clouds at 0.01 bar (bottom).} 
	\centering 
	\label{fig:spectra}
	\end{figure}

	The line lists were taken from \citet{polyansky_h2o}, \citet{rothman_co2}, and \citet{rothman_so2} for H$_2$O, CO$_2$ and SO$_2$ respectively. For each of these we showcase both a cloud-free atmosphere and one with an opaque grey cloud layer at 0.01 bar which mutes the features seen. These are simplified circumstances but highlight the type and strength of features that could be seen through transmission spectroscopy of this planet. We model these spectra over the range 0.4 - 15 $\mu$m and our synthetic spectra are shown in Fig~\ref{fig:spectra}.
	
	We are aware that the presence of nitrogen-rich atmospheres on warm or hot super-Earths is a subject of much dispute. For example, nitrogen atmospheres have been proposed on some molten super-Earths such as 55~Cancri~e \citep[e.g.][]{Angelo2017,Hammond2017,Miguel2019,Zilinskas2020}. However, there are problems with this model, such as for example the inability of nitrogen to trigger efficient cooling, resulting in eventual hydrodynamical losses \citep[e.g.][]{Tian2008,Lichtenegger2010,Airapetian2017,Johnstone2019}. Furthermore, some argue that exogenous and endogenous processes could trigger the deposition of atmospheric $\rm N_{2}$ \citep[e.g.][]{Navarro2001,Parkos2018,Lammer2019}. Nevertheless, it is not the objective of the present study to challenge or support the possibility of $\rm N_{2}$ atmospheres on warm or hot super-Earths and so we include nitrogen within our synthetic spectrum as a proof of concept.

	In any case, given the brightness of the host star and the significant features over the wavelength ranges covered by Twinkle (0.5-4.5 $\mu m$, \cite{twinkle}), Ariel (0.5-7.8 $\mu m$, \cite{tinetti_ariel}), and the James Webb Space Telescope (JWST, 0.6-12 $\mu$m), the atmospheric composition of GJ~357~b could be constrained using these instruments which would provide important information for the interior structure and geochemistry of the planet. Additionally, the G141 grism of the Hubble Space Telescope's Wide Field Camera 3, which covers 1.1-1.7 $\mu m$, may be sensitive enough to infer the presence of water. A sufficiently high abundance of water could indicate a large-scale planetary migration from beyond the ice-line to its current location or a late planetesimal bombardment. Conversely, chemical species such as $\rm CO_{2}$ and $\rm SO_{2}$ would indicate an Earth-like composition with a volcanically active past.
	
	We acknowledge that we have simulated a relatively ideal case for the transmission profile which neglects the possible presence of complex hazes that might be expected to be present in the atmosphere of a volcanic planet. We also do not account for chemical reactions between the different species. There are many processes that could alter the chemical composition of the atmosphere. One such process is XUV-irradiation which could substantially affect the stability of molecules. The XUV irradiation that GJ~357~b receives is approximately equal to eight times that of Venus but it also has a stronger gravitational force. In addition the presence of a geomagnetic field could further mitigate the XUV-induced photoevaporation \citep{Owen2019}. Furthermore, even in the absence of a geomagnetic field, the effects of ionising photons could be suppressed if an ionosphere forms that interacts with the escaping ions. This mechanism is thought to take place on Venus \citep{Zhang2012} and it may explain how some volatiles such as water vapour are still retained (albeit in small abundances). We understand that some species can dissociate into their elemental constituents when exposed to XUV irradiation, which would ultimately affect the bulk atmospheric composition. Water for example would form oxygen and hydrogen and whilst oxygen may be retained due to its heavier mass, the hydrogen could be lost. However, due to this being a multivariate problem we still consider water but in a low abundance ($1\%$) in our synthetic spectra as it could be present. To summarise, while the actual spectrum of the planet may display a higher complexity because of the simplifying assumptions in our synthetic spectra, our model highlights atmospheric spectral features that might plausibly be detected in an eventual observational study of GJ\,357\,b. A thorough investigation of the ability of current and upcoming facilities to disentangle potential atmospheric scenarios is left for future work.

	\section{The robustness of the planetary density}\label{sect:stellarparams}

	Our model for the secondary atmosphere of GJ\,357\,b is based on the bulk density of the planet that is derived from its mass and radius. In addition to the typical uncertainties associated with radial-velocity and photometric measurements used to derive the ratios between planetary and stellar masses and radii, one of the main sources of systematic error are the estimates of the stellar mass and radius. This is because the planetary properties are all measured as a function of the stellar quantities. In order to verify the robustness of our results against errors due to the methods used to derive the stellar properties, we recorded the changes in the planetary density for various estimates of the stellar parameters. In particular, we consider the stellar mass ($M_*$) and radius ($R_*$) derived in the following distinct ways.
	
	First, we followed the methodology from  \citet[][hereafter
	MA15]{2015A&A...577A.132M}\footnote{https://github.com/jesusmaldonadoprado/mdslines} which is based on the use of optical
	high-resolution spectra from HARPS. A total of $53$ spectra were downloaded from the ESO archive\footnote{http://archive.eso.org/wdb/wdb/adp/\\phase3\_spectral/form?phase3\_collection=HARPS}, which were collected from radial-velocity measurements and co-added into one single spectrum. Initially, the effective temperature and metallicity of the star are computed from ratios of pseudo equivalent widths of
	spectral features. The effective temperature is calibrated using stars with interferometric estimates of their radii and is in the revised scale by \cite{2013ApJ...779..188M}. The stellar metallicities provided by MA15 are based on the photometric M$_{\rm K}$-[Fe/H] relationship by \cite{2012A&A...538A..25N}.
	Derived values of T$_{\rm eff}$ and [Fe/H] are 3461 $\pm$ 68 K and -0.14 $\pm$ 0.09 dex, respectively. MA15 also provides empirical calibrations to derive the stellar evolutionary parameters as a function of the stellar effective temperature and the metallicity. The mass scale in MA15 is based on the NIR photometric calibration of \cite{1993AJ....106..773H} and has typical uncertainties of the order of 13\%. MA15 derive their own stellar mass$-$radius relationship using stars with known interferometric radius and low-mass eclipsing binaries. Typical uncertainties in the radius are of the order of 12\%. 
	
	Secondly, we obtain $R_{\rm *}$ and $M_{\rm *}$ from the empirical   radius$-$magnitude and mass$-$magnitude relationships of \citet[][their Eqs.~4 and~10]{2015ApJ...804...64M}. Here we used the {\it Gaia} parallax to obtain the absolute $K_{\rm s}$ band magnitude from the 2\,MASS measurement (see Sect.~\ref{subsect:xrays_analysis}). This method was applied in the activity$-$rotation$-$age study of M dwarfs of \citet{Magaudda2020}. Therefore, we refer to these results as MSC\,20. 

	We also consider the published values from \citet[][hereafter SC19]{2019A&A...625A..68S}. Their analysis comprises the following steps: First, bolometric luminosities ($L_{\rm bol}$) are determined by integration of the available photometry. Then, effective temperature ($T_{\rm eff}$) and stellar metallicity ([Fe/H]) are obtained by fitting the optical CARMENES spectra to a set of PHOENIX-ACES synthetic spectra. Surface gravity, $\log g$, is fixed by a T$_{\rm eff}-\log g$ relation from theoretical $5$\,Gyr isochrones \citep{2018A&A...615A...6P}. The stellar radius is computed from $T_{\rm eff}$ and $L_{\rm bol}$ using the Stefan-Boltzmann law. Finally, the authors derive the stellar mass using their own mass$-$radius relationship calibrated with a sample of 55 eclipsing M dwarf binaries (their Eqn.~6). In the following, we denote these values as ($M_{\rm M-R}$, $R_{\rm SB}$).
	
	In addition to these estimates, SC19 compared their values of stellar mass and radius with those obtained from another three methods: (i) Spectroscopic mass ($M_{\log g}$) derived from the values of $\log{g}$, and radius computed before. (ii) Photometric mass ($M_{\rm M-Ks}$) computed using the mass$-$2MASS $K_{\rm s}$ magnitude relation by \citet{2019ApJ...871...63M}. (iii) Mass and radius based on the PARSEC evolutionary models ($M_{\rm PAR}$, $R_{\rm PAR}$) computed using a Bayesian approach applied to the PARSEC stellar library as in \cite{2018MNRAS.479.1953D}.

\begin{figure}[htb] 
		\includegraphics[scale=0.5,center]{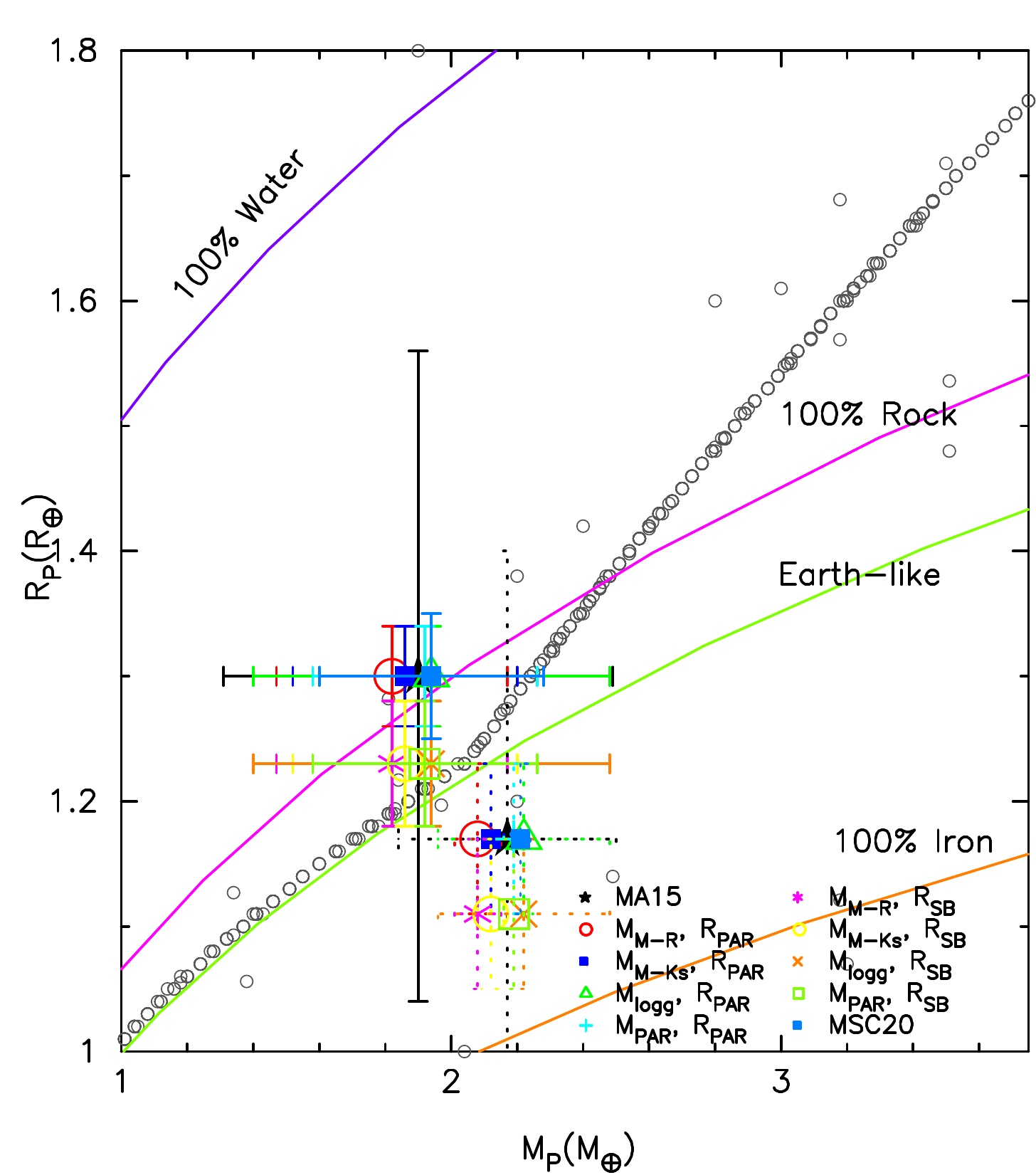}
		\caption{Position of GJ~357~b in the radius-versus-mass diagram.
		Different colours indicate the planetary parameters derived using the
		different datasets of stellar masses and radii.
		Continuous lines indicate the estimates based on the Keplerian amplitude and planet-to-star radius ratio by
		\cite{Luque2019} while dotted lines show the results based on the analysis by \cite{Jenkins2019}. A few exoplanets from NASA's exoplanets archive are overplotted as grey circles. Different planetary models \citep[from ][]{Zeng2013,Zeng2016} are also shown for comparison.}
		
		\label{diagrama_hr} 
		\end{figure}

	Table~\ref{stellar_parameters_effects} provides the stellar mass and radius for each analysis method. The most important difference between the methods arises from the use of different mass$-$luminosity and mass$-$radius relationships. In order to test the effect of the stellar mass on the
	determination of the planetary mass we made use of the published values of the radial-velocity semi-amplitude, $K_* = (1.52 \pm 0.25)~\rm ms^{\rm -1}$ and orbital period, $P = 3.93072^{\rm + 0.00008}_{\rm - 0.00006}~\rm days$ \citep{Luque2019}. We then derived for each estimate of the stellar radius a value for the radius of GJ\,357\,b using the relation $R_{\rm P}/R_{\star}$ = 0.0331 $\pm$ 0.0009 \citep{Luque2019}. Using the $R_{\rm P}$ and $M_{\rm P}$ values derived that way, an estimation of the planetary density was performed. The results are provided in Table~\ref{stellar_parameters_effects}, while Fig.~\ref{diagrama_hr} shows the position of GJ~357~b in the radius-versus-mass diagram according to our different estimates. 
	
	We find that the mass and radius values for GJ\,357\,b based on a given method to derive the stellar parameters of the host star exhibit considerable uncertainties. However, our comprehensive comparison of different ways to calculate the stellar parameters of GJ\,357 shows that all of these methods yield consistent results in terms of the planet mass and radius. Therefore, the stellar parameters introduce negligible effects on our model for the atmospheric mass and composition of GJ\,357\,b.
	
	As mentioned previously, another factor that might influence the estimation of the planetary density is the analysis used on the computation of the Keplerian amplitude and the ratio $R_{\rm P}/R_{\star}$ due to the GJ357 b planet. Indeed, recently \citet[][Table~4]{Jenkins2019} provided a slightly larger Keplerian amplitude $K_* = (1.7372^{\rm + 0.0054}_{\rm - 0.0007} ms^{\rm -1}) $ and a smaller star-to-planet radius ratio, $R_{\rm P}/R_{\star}$ = 0.02981 $\pm$ 0.0013. This translates into higher masses and densities for GJ357 b, which are also listed in Table~\ref{stellar_parameters_effects} and shown in Figs.~\ref{fig:mass_radius} and \ref{diagrama_hr}. The slightly different planetary parameters between \cite{Jenkins2019} and \cite{Luque2019} might be related to the use of different datasets. While \cite{Jenkins2019} use HARPS, UVES, and HIRES data, the analysis done in \cite{Luque2019} also includes CARMENES and PSF data. If the planetary densities based on the \cite{Jenkins2019} results are considered, the composition of the planet would be consistent with a telluric planet that has a core $\simeq 60~\%$ \citep{Zeng2016} of the total mass or with the icy remnant core of a planet which once hosted a large primordial atmosphere \citep{Mocquet2014,Modirrousta2020b}. However, as explained previously, the uncertainties in the measured values allow for a more typical rocky composition.
	
\begin{table*}
\centering
\caption{Planetary parameters of GJ~357~b derived from the different datasets of stellar parameters of GJ~357.}
\label{stellar_parameters_effects}
\begin{scriptsize}
\begin{tabular}{lcccccccc}
\hline
Dataset$^{\dag}$ & M$_{\star}$    &  R$_{\star}$   &  m$_{\rm P}$[LU] & R$_{\rm P}$[LU] & $\rho_{\rm P}$[LU] & m$_{\rm P}$[JE] & R$_{\rm P}$[JE] &  $\rho_{\rm P}$[JE]   \\
                 & (M$_{\odot}$)  &  (R$_{\odot}$) & (M$_{\oplus}$)   & R$_{\oplus}$)   &   (gcm$^{\rm -3}$ &(M$_{\oplus}$)   & R$_{\oplus}$)   &  (gcm$^{\rm -3}$)    \\
\hline
 Ma15                                &       0.36   $\pm$       0.08  &       0.36    $\pm$       0.07  &         1.90 $\pm$         0.59&         1.30 $\pm$         0.26&         4.76 $\pm$         3.18&    2.17 $\pm$         0.33&         1.17 $\pm$         0.23&         7.45 $\pm$         4.59 \\
 SC19: M$_{\rm M-R}$, R$_{\rm PAR}$  &       0.3368 $\pm$       0.0150&       0.3601  $\pm$       0.007 &         1.82 $\pm$         0.35&         1.30 $\pm$         0.04&         4.55 $\pm$         0.99&    2.08 $\pm$         0.07&         1.17 $\pm$         0.06&         7.12 $\pm$         1.05 \\
 SC19: M$_{\rm M-Ks}$, R$_{\rm PAR}$ &       0.3477 $\pm$       0.0084&       0.3601  $\pm$       0.007 &         1.86 $\pm$         0.34&         1.30 $\pm$         0.04&         4.65 $\pm$         0.96&    2.12 $\pm$         0.04&         1.17 $\pm$         0.06&         7.27 $\pm$         1.05 \\
 SC19: M$_{\log g}$, R$_{\rm PAR}$   &       0.3716 $\pm$       0.064 &       0.3601  $\pm$       0.007 &         1.94 $\pm$         0.54&         1.30 $\pm$         0.04&         4.86 $\pm$         1.44&    2.22 $\pm$         0.26&          1.17 $\pm$         0.06&         7.60 $\pm$         1.41 \\
 SC19: M$_{\rm PAR}$, R$_{\rm PAR}$  &       0.3653 $\pm$       0.0071&       0.3601  $\pm$       0.007 &         1.92 $\pm$         0.34&         1.30 $\pm$         0.04&         4.80 $\pm$         0.98&    2.19 $\pm$         0.03&         1.17 $\pm$         0.06&         7.52 $\pm$         1.09 \\
 SC19: M$_{\rm M-R}$, R$_{\rm SB}$   &       0.3368 $\pm$       0.015 &       0.3419  $\pm$       0.011 &         1.82 $\pm$         0.35&         1.23 $\pm$         0.05&         5.32 $\pm$         1.23&    2.08 $\pm$         0.07&          1.11 $\pm$         0.06&         8.32 $\pm$         1.38 \\
 SC19: M$_{\rm M-Ks}$, R$_{\rm SB}$  &       0.3477 $\pm$       0.0084&       0.3419  $\pm$       0.011 &         1.86 $\pm$         0.34&         1.23 $\pm$         0.05&         5.43 $\pm$         1.20&    2.12 $\pm$         0.04&         1.11 $\pm$         0.06&         8.50 $\pm$         1.39 \\
 SC19: M$_{\log g}$, R$_{\rm SB}$    &       0.3716 $\pm$       0.064 &       0.3419  $\pm$       0.011 &         1.94 $\pm$         0.54&         1.23 $\pm$         0.05&         5.68 $\pm$         1.74&    2.22 $\pm$         0.26&          1.11 $\pm$         0.06&         8.88 $\pm$         1.78 \\
 SC19: M$_{\rm PAR}$, R$_{\rm SB}$   &       0.3653 $\pm$       0.0071 &       0.3419 $\pm$       0.011 &         1.92 $\pm$         0.34&         1.23 $\pm$         0.05&         5.61 $\pm$         1.22&    2.19 $\pm$         0.03&         1.11 $\pm$         0.06&         8.78 $\pm$         1.44 \\
 MSC20 method                        &       0.37   $\pm$        0.007&          0.36 $\pm$       0.01 &          1.94 $\pm$         0.34&         1.30 $\pm$         0.05&         4.85 $\pm$         1.03&    2.21 $\pm$         0.03&         1.17 $\pm$         0.06&         7.59 $\pm$         1.19 \\
\hline
\end{tabular}
\tablefoot{$^{\dag}$ MA15 \citep{2015A&A...577A.132M}; SC19 \citep{2019A&A...625A..68S}: M$_{\rm M-R}$ mass from mass-radius relationship; M$_{\rm M-Ks}$ photometric mass; M$_{\log g}$ spectroscopic mass; M$_{\rm PAR}$ PARSEC-based mass; R$_{\rm PAR}$ PARSEC-based radius; R$_{\rm SB}$ radius from the Stefan-Boltzmann's law; IRPH: Mass and radius from the relationships by \cite{2015ApJ...804...64M}. m$_{\rm P}$[LU],   R$_{\rm P}$[LU],  $\rho_{\rm P}$[LU] refer to the planetary parameters obtained using the planetary keplerian amplitude and the planet-to-star radius ratio by \cite{Luque2019}, while m$_{\rm P}$[JE], R$_{\rm P}$[JE], $\rho_{\rm P}$[JE] estimates are based on the analysis by \cite{Jenkins2019}.  }
\end{scriptsize}
\end{table*}

		\section{Summary and conclusions}\label{sect:discussion}

		Our analysis of GJ\,357\,b and its host star has led to the following results and conclusions:  
		
		From a recent XMM-Newton observation we derived an extremely low X-ray luminosity for GJ\,357 ($\log{L_{\rm x}\,{\rm [erg/s]}} = 25.7$ in the $0.1-2.4$\,keV {\em ROSAT} band). 
		When compared to the $L_{\rm x}$ of most similar-mass M dwarfs with known ages and to different $L_{\rm x}- \rm age$ laws for M dwarfs \citep{Magaudda2020, Penz2008(1)} this low X-ray activity indicates that GJ\,357 is at least $5$\,Gyr old and possibly significantly older. Under the assumption that the star and planet formed at a similar time, we can assume $5$~Gyr to be a conservative estimate for the age of the planet system. 
		
        Using the X-ray luminosity of GJ~357 with our above estimate for the age of the system and the empirical $L_{\rm x}- \rm age$ relation from \cite{Penz2008(1)} we performed a backwards reconstruction of GJ~357~b's primordial atmospheric mass. We find a theoretically maximum envelope mass of $\sim 38 \rm M_{\oplus}$. However, it is unlikely that GJ~357~b accreted an envelope this massive, as its small central mass would have hindered its ability to collect gas. Conversely, if one adopts the mass and radius measurements from \citet{Jenkins2019} instead of those from \citet{Luque2019}, then GJ~357~b has a density consistent with a compressed remnant core \citep{Mocquet2014}. Its high density could be suggestive of an initially large atmospheric mass.
		
		Taking account of the known parameters of the host star and the estimated planet mass and radius, we have produced three different synthetic IR spectra representing the potential secondary atmosphere of GJ\,357\,b. Our test cases comprise (a) $100\%~\rm CO_{2}$, (b) $100\%~\rm SO_{2}$, (c) $\rm 75\%~ N_{2}$ with  $\rm 24\%~ CO_{2}$ and $\rm 1\%~ H_{2}O$. The actual  atmospheric content of this planet should be accessible to upcoming space missions, such as Ariel, JWST, and Twinkle, which will allow us to test our predictions. Finally, a detailed analysis of the uncertainties in the stellar and planetary parameters shows that despite the uncertainties being considerable, they do not have significant effects on our results.

\begin{acknowledgements} We acknowledge the support of the Ariel ASI-INAF agreement n.2018-22-HH.0. EM is supported by the Bundesministerium für Wirtschaft und Energie through the Deutsches Zentrum für Luft- und Raumfahrt e.V. (DLR) under grant number FKZ 50 OR 1808. JSF acknowledges support by the Spanish MICINN grant AYA2016-79425-C3-2-P. MG was supported by the Austrian Science Fund (FWF) project S116 “Pathways to Habitability: From Disk to Active Stars, Planets and Life” and the related  subproject S11604-N16. BE is funded by the STFC grant ST/T001836/1.
We thank L. Mugnai for his useful suggestions. This work makes use of observations obtained with XMM-Newton, an ESA science mission with instruments and contributions directly funded by ESA Member States and NASA. We thank the anonymous referee for their useful comments.
\end{acknowledgements}

\bibliographystyle{aa} \bibliography{bibliography.bib}

\end{document}